\documentclass[twocolumn,preprintnumbers,superscriptaddress,showpacs,showkeys,twoside]{revtex4}

\usepackage{color}
\usepackage{graphicx}
\usepackage{dcolumn}
\usepackage{bm}

\renewcommand\textfraction 0
\renewcommand\topfraction 1
\renewcommand\bottomfraction 1
\preprint{Submitted to Applied Surface Science (EMRS-2004 \# N-I.3) }

\begin{document} 

\hyphenation{ext-ra-po-la-ting na-no-clus-ter na-no-clus-ter na-no-se-cond fem-to-se-cond}


\title{Production of Gas Phase Zinc Oxide Nanoclusters by Pulsed Laser Ablation}
\author{I. Ozerov}
\email{ozerov@crmcn.univ-mrs.fr}
\affiliation{CRMC-N, UPR 7251 CNRS, Universit\'{e} de la M\'{e}diterran\'{e}e, Case 901, 13288 Marseille Cedex 9, France.}

\author{A. V. Bulgakov }
\affiliation{Institute of Thermophysics SB RAS, 1 Lavrentyev Ave., 630090 
Novosibirsk, Russia.}

\author{D. K. Nelson}
\affiliation{A.F. Ioffe Physico-Technical Institute, Polytechnicheskaya 26, 194021  St.-Petersburg, Russia.}

\author{R. Castell}
\affiliation{Universidad Sim\'{o}n Bol\'{\i}var, 1080 Caracas, Venezuela.}

\author{W. Marine}
\email{marine@crmcn.univ-mrs.fr}
\affiliation{CRMC-N, UPR 7251 CNRS, Universit\'{e} de la M\'{e}diterran\'{e}e, Case 901, 13288 Marseille Cedex 9, France.}

\date{\today}
\begin{abstract}

We present experimental results on the photoluminescence (PL) of 
gas-suspended zinc oxide nanoclusters prepared during ablation of 
sintered ZnO targets by a pulsed ArF laser in the presence of oxygen ambient gas. The PL 
spectra in the UV spectral region correspond to the exciton recombination in 
the nanoclusters which are crystallized and cooled down to the temperature of the ambient gas in the ablation 
chamber. The time evolution of the spectra as well as their dependence on 
the ambient gas pressure are discussed.



\end{abstract}
\maketitle


\section{Introduction}

Nanocrystalline wide-gap semiconductors have attracted a great interest due to their
importance in fundamental research and potential technological applications.
In particular, zinc oxide is considered as one of the most promising materials for realization of optoelectronic devices like UV light emitting diodes and lasers
\cite{Bagnall,Tang,Choopun,Cao}.
Several physical and chemical methods like molecular-beam epitaxy \cite{Bagnall}, chemical vapor deposition \cite{Haga}, sputtering \cite{Yoon} have been successfully applied for ZnO film synthesis.
The technique of nanosecond \cite{Tang,Choopun,Cao,Nakamura,ApSuSc2003,Hartanto} and femtosecond \cite{Millon} pulsed laser ablation (PLA) has been also widely used to produce both epitaxial and nanostructured ZnO films of high optical quality. 
The synthesis of nanocrystalline films is usually performed  in an ambient gas
whose pressure should be high enough, firstly,
to confine the plasma created after laser ablation in order to favor
the collisions and chemical reactions in the laser induced plume and,
secondly, to induce a rapid cooling of the plasma to favor the condensation of the nanoclusters
in gas phase \cite{Strategy}. Previous experimental optical studies of the laser-induced plumes \cite{Kawaguchi2002,Claeyssens2002} in vacuum and in ambient gas did not show the evidence for ZnO cluster formation.
Recently, we have observed small nanoclusters with composition Zn$_{n}$O$_{m}$ (with $n, m$ up to 15) by time-of-flight mass-spectrometry \cite{Bulgakov} in vacuum and in gas at low 
pressures. The formation of these atomic clusters in laser induced plasma 
can be considered as the first step in the synthesis of clusters with sizes 
of several nanometers which are usually deposited on a substrate 
individually or in form of a thin film \cite{Hartanto,ApSuSc2003,Strategy}.
Despite of numerous reports on laser deposition of nanostructured films, few data are
available on the origin of the nanoclusters and on their aggregate state during the 
transport to the substrate. However, this 
information is crucial to control and, consequently, optimize the deposition conditions of nanostructured films. 
In this paper we show that ZnO nanoclusters are well formed in the gas phase and we present the results on \textit{in situ} photoluminescence from the clusters during their synthesis by PLA in the oxygen ambient.

\section{Experimental}

The nanoclusters were prepared by pulsed laser ablation of a sintered ZnO target. The target was placed inside a stainless steel chamber on a rotating holder. After evacuating the chamber down to a pressure of about $2\times 10^{-7}$ mbar by a turbomolecular pump, a continuous slow flux of oxygen, used as ambient gas, was introduced into the chamber at a controlled pressure ranging from 2 to 8 mbar. An ArF excimer laser (Lambda Physik LPX 205) operating at a wavelength of 193 nm with pulse duration of 15 ns (FWHM) was used for the ablation. The laser beam was focused onto the target with an incident angle of 45$^{\circ}$ and the fluence was about 3.5 J/cm$^{2}$. The choice of these conditions was guided by our previous results on the deposition of nanocrystalline films \cite{ApSuSc2003}. A typical ablation run time was about 5 minutes, that corresponds to 900 laser shots at a repetition rate of 3 Hz. After the end of the ablation procedure, the photoluminescence (PL) spectra of the clusters have been taken \textit{in situ}. The photoluminescence was excited either by the non-focused beam of the ArF laser or by the 3rd harmonic of a Ti:sapphire femtosecond laser (Spectra Physics, $\lambda = $ 266 nm, $\tau = $ 100 fs) and analyzed by a double monochromator (Jobin-Yvon, DH-10) provided with a photomultiplier and a boxcar integrator (Stanford SR-250). The system was synchronized using an external delay generator (Stanford DG-535). The boxcar was started before the laser shot and the gate was 50 ns in order to overcome the timing jitter problem. Gated intensified CCD camera (Andor DH-734) was used for rapid imaging.

\section{Results and discussions}

\begin{figure}[!]
\resizebox{7.5cm}{!}{\includegraphics{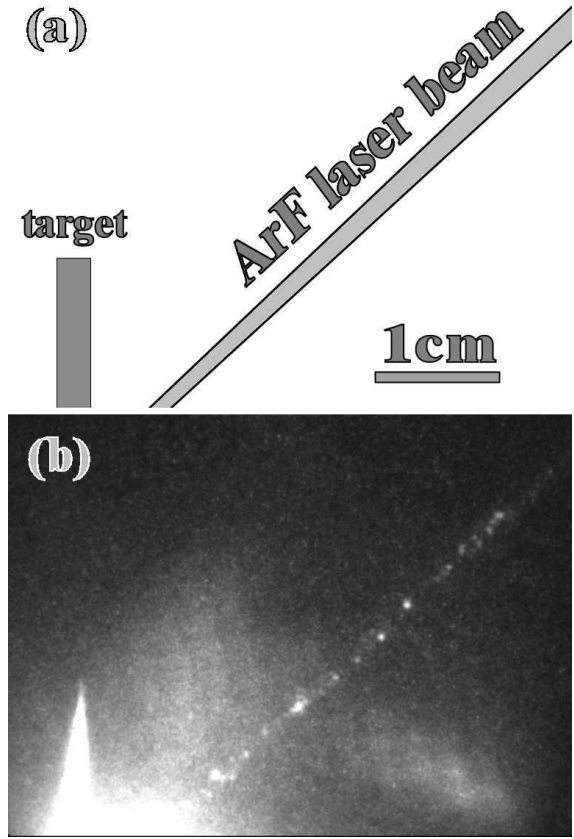}} \caption{\label{iCCDimage}
(a) Sketch of the ablation geometry. (b) I-CCD camera image 
of ZnO nanoclusters excited by incident ArF laser beam.}
\end{figure}

\begin{figure}[!]
\resizebox{7.5cm}{!}{\includegraphics{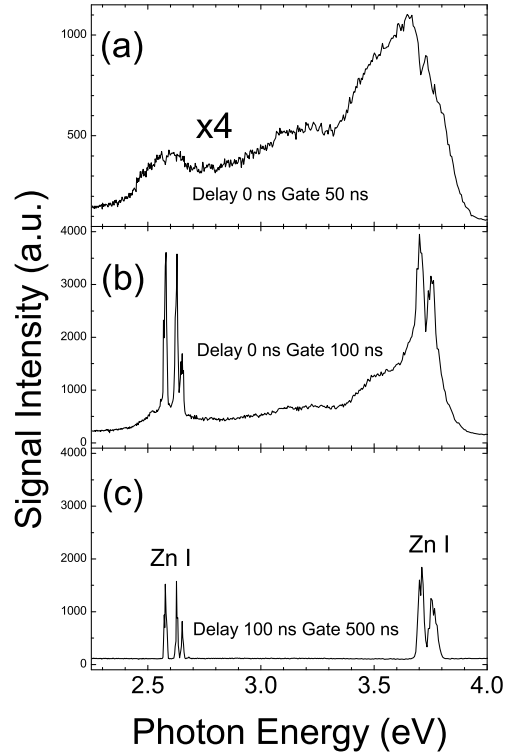}}
\caption{\label{fragmentation}
Fragmentation of gas-suspended ZnO nanoclusters excited by ArF laser
(fluence $ \sim 200 $ mJ/cm$^{2}$): (a) Boxcar delay 0 ns, gate 50 ns; (b) delay 0 ns, gate 100 ns; (c) delay 100 ns, gate 500 ns. The oxygen pressure was 4 mbar. The atomic lines of Zn are indicated in the figure. The observation point was at a distance of $\sim 5 $ cm from the target.}
\end{figure}

\begin{figure}[!]
\resizebox{7.5cm}{!}{\includegraphics{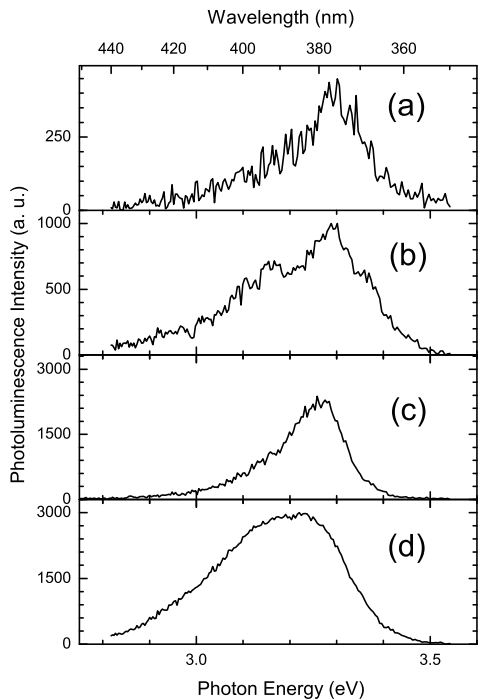}}
\caption{\label{plpressure}
PL spectra of gas-suspended ZnO nanoclusters excited by ArF laser
(fluence $ 25 - 35 $ mJ/cm$^{2}$) 
taken just after the ablation run for different ambient gas pressures: (a) 2 
mbar; (b) 6 mbar. PL spectra of a ZnO nanocrystalline film 
excited at different laser fluences: (c) $\sim 30 $ mJ/cm$^{2}$ and (d) $\sim 90 $  
mJ/cm$^{2}$.}
\end{figure}

\begin{figure}[!]
\resizebox{7.5cm}{!}{\includegraphics{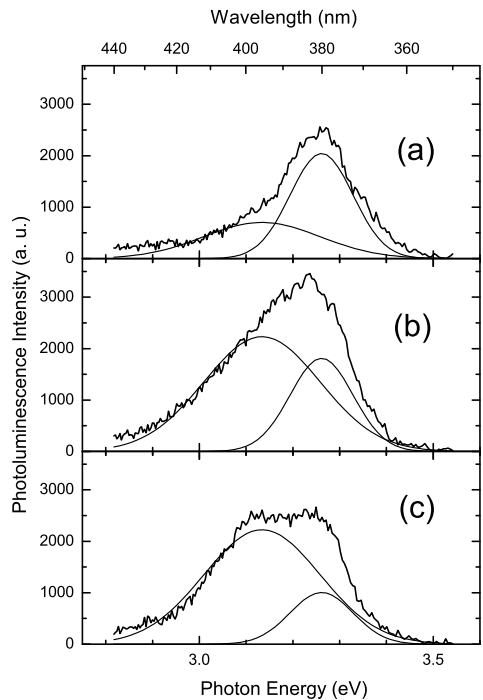}} \caption{\label{pl8mbar}
Evolution of the PL spectra of gas-suspended ZnO nanoclusters. The oxygen pressure was 8 mbar.
(a) 1st pass (0 minutes);
(b) 2nd pass (delay 3 min); (c) 3rd pass (delay 6 min). The fluence of ArF laser was $\sim 30 $  mJ/cm$^{2}$. Thin lines show the results of fitting by two Gaussians.}
\end{figure}

\begin{figure}[!]
\resizebox{7.5cm}{!}{\includegraphics{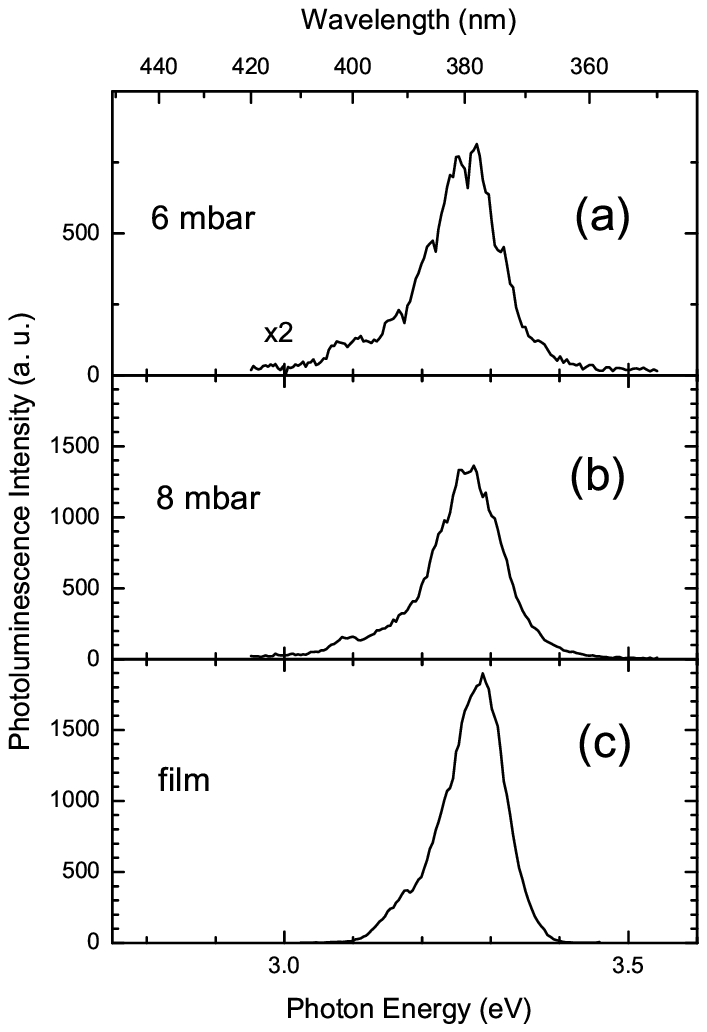}} \caption{\label{plfemto}
PL spectra of gas-suspended ZnO nanoclusters for different ambient 
gas pressures: (a) 6 mbar; (b) 8 mbar. PL spectrum of a ZnO 
nanocrystalline film (c). The PL was excited by 3$^{rd}$ harmonics of the 
femtosecond laser $\lambda = 266$ nm. The fluence was about 300 $\mu$J/cm$^{2}$}
\end{figure}

During the experiments on the deposition of nanocrystalline films 
 we have observed the presence of ZnO aggregates suspended in 
the gas by an intense luminescence along the path of the incident ablating 
ArF laser beam as illustrated in Fig. \ref{iCCDimage}. Figure \ref{iCCDimage}a shows a sketch of the ablation system. The target is situated at the left lower part of the image and the ablating 
laser beam hits the target at a 45$^{\circ}$ incident angle. Figure \ref{iCCDimage}b presents 
an iCCD camera image which was taken simultaneously with the ablation laser 
pulse. One can clearly see the luminescent particles in the path of the ablating
laser beam. The clusters have been accumulated in the gas phase from 
precedent laser shots and slowly displaced within the ambient gas. The 
luminescence is observed after several ablation shots needed for 
accumulation of a detectable quantity of the clusters in the chamber. The clusters are the product of a complex process including gas phase condensation and laser-induced fragmentation.
At the observation point situated far from the focal plane, the fluence of 
the focused ArF laser beam was about 200 mJ/cm$^{2}$. This fluence is fairly high to induce the evaporation of atomic species from the nano-aggregates and their 
fragmentation \cite{JdePhys,Boufendi}. The corresponding luminescence spectra are shown in Fig. \ref{fragmentation}. The spectrum in Fig. \ref{fragmentation}a was taken during the laser pulse. The spectrum presents several superposed bands originated from both emission of ZnO nanoparticles and the continuum due to the bremsstrahlung emission \cite{Claeyssens2002}. At a time delay of about 50 ns relatively to the laser pulse, narrow emission lines of atomic zinc appeared in the spectrum together with the continuum emission (Fig. \ref{fragmentation}b). These Zn atoms are desorbed or evaporated from clusters. At longer delays the spectrum consists of only zinc atomic lines (Fig. \ref{fragmentation}c).

In order to prevent the fragmentation of the nanoclusters during the PL measurements, we have accumulated the clusters in gas phase and
studied their PL under the excitation by additional laser beams after the  
end of the ablation run.  Figure \ref{plpressure} shows the PL spectra of the gas-suspended ZnO nano-aggregates
prepared in the oxygen ambient gas at different pressures.
The spectra have been taken just after the ablation run
and have been excited by a non-focused ArF laser beam.
The PL intensity increases with ambient gas pressure indicating the increase in the amount of clusters flying in the chamber.
The spectra consist of a complex band of an asymmetric shape maximized
at a photon energy of 3.30 eV. The band have large tails expanding on both
high and low energy sides of the spectrum.

We attribute this photoluminescence to the ZnO nano-aggregates already 
crystallized and cooled by collisions with molecules of the ambient gas. 
The spectral position of the peak corresponds to the recombination of excitons in the  
clusters \cite{ApSuSc2003,VanDijken}. We note that the defect-related band observed usually in ZnO at photon energies of $ 2 - 2.5 $ eV \cite{ApSuSc2003} is absent.
It is interesting to compare the PL of gas-suspended clusters with 
the PL obtained from a nanocrystalline film under similar excitation of 30 mJ/cm$^{2}$
(Fig. \ref{plpressure}c). The 
excitonic transition in the film is observed at the photon energy of 3.28 eV and the 
spectrum is narrower than that observed for gas-suspended nanoclusters.
Also, the high energy tail is strongly reduced. The sizes of nanoclusters deposited
in form of the film were about $8 - 10$ nm \cite{ApSuSc2003}.
The maximum position of the exciton emission band from gas-suspended nanoclusters
is shifted to about 20 meV higher photon energies compared to the PL from the films.
Both features, the blue shift of the peak and the high energy tail of the PL spectra indicate the possibility of the quantum confinement in small nanoclusters which are present in the gas phase.
It means that a large amount of gas-suspended clusters have the sizes comparable with the exciton Bohr radius which is 1.8 nm in ZnO \cite{VanDijken}.

All the spectra obtained from the nanoclusters under the nanosecond laser excitation have
a low-energy tail of the exciton emission band comparable to that observed from a film under high excitation (figure \ref{plpressure}d). This low-energy tail of the PL spectra of gas-suspended clusters exhibits a slow evolution in minute time scale indicating the cluster coalescence dynamics.
Figure \ref{pl8mbar} shows the evolution of the PL spectra of ZnO nanoclusters prepared 
in 8 mbar of oxygen. The first spectrum (Fig. \ref{pl8mbar}a) was taken just after the 
ablation run. We should note the higher integral intensity of the spectrum compared to those obtained at lower ambient gas pressure (Fig. \ref{plpressure}a,b). This fact as well as stronger scattering of a continuous laser beam which we used to test for the presence of clusters indicate that the concentration of the nanoclusters increases with ambient gas pressure. At the same time, the shape of the PL peak observed in Fig. \ref{pl8mbar}a is similar to that observed for lower gas pressures. 
The second (Fig. \ref{pl8mbar}b) and the third (Fig. \ref{pl8mbar}c) spectra were taken 3 
and 6 minutes after the first one, respectively.
The shape of the PL spectra indicates that they are originated from two different transitions. In order to separate the contributions of these two overlapped bands, the spectra have been deconvoluted by two Gaussian functions. The maxima of the bands are centered at 3.26 and 3.13 eV with FWHM of 0.13 and 0.25 eV respectively. The first band decreases in intensity with time while the second, low energetic one, increases from the first to the second spectra and then remains constant.
The cluster PL signal taken at long time delays (Fig. \ref{pl8mbar}b,c) is similar in shape to that observed from a film excited with high fluence (Fig. \ref{plpressure}d). 
These results provide evidence for the growth and/or coalescence of the nanoclusters. It is known that the efficiency of light absorption by nanoparticles increases with the particle size (see \cite{Boufendi} and references therein). Thus, when the clusters grow up they absorb more photons and more electron-hole pairs are created leading to the increase in the PL intensity.

Two possible mechanisms lead to the increase of the low energy part of the PL spectra. First, if high intensity laser excitation creates the electron-hole plasma, the emission peak is shifted because of the bandgap energy (E$_{g}$) renormalization. The red shift of the emission peak from ZnO epitaxial films with increase of the excitation intensity has been reported under femtosecond excitation \cite{Yamamoto}.
Second, the excitation laser beam can heat the nanoclusters resulting in the decrease of the optical bandgap and in the red shift of the interband transitions. By extrapolating the temperature dependence of bulk ZnO optical bandgap \cite{Wang}, we found that the gap energy decrease down to 3.13 eV corresponds to heating up to a temperature of about $ 550 - 600 $ K. 

When photons with energy of 6.4 eV are absorbed in a material with E$_{g} = 3.3 $ eV hot carriers are created. The thermalization of these hot carriers via electron-phonon interaction leads to the cluster heating on a time scale of several picoseconds followed by radiative recombination. In the case of non-radiative recombination of the free carriers, all the photon energy is directly converted to heat.
We estimate the increase in temperature of a ZnO cluster with size of about 1.8 nm 
after absorption of one photon of 6.4 eV energy to be about 20 K.
The number of photons absorbed by a particle irradiated at the fluence of 30 mJ/cm$^{2}$ used in our experiments is  
estimated to be about 15 taking into account the nanocluster absorption efficiency \cite{Boufendi}. This number gives a relatively good approximation of the cluster temperature of $ 550 - 600 $ K
estimated from the PL spectra.

In order to confirm the mechanism responsible for the red tail of the emission peak from ZnO nanoclusters we have excited the PL by a femtosecond laser at the fluence of about 300 $\mu$J/cm$^{2}$ giving a slightly higher number of excited carriers than the nanosecond laser at a fluence of 30 mJ/cm$^{2}$. The concentration of free carriers ($N$)  has been estimated using the rate equation:
\begin{equation}
\label{eq:rate}
\frac{dN}{dt} = G - \frac{N}{\tau}, ~~~~G = \frac{(1-R)\alpha F}{\tau_{laser} h\nu},
\end{equation}
\noindent 
where  $G$ is the carrier generation term, $R$ is a reflectivity, $\alpha$ is the absorption coefficient, $F$, $\tau_{laser}$ and $h\nu$ are laser fluence, pulse duration and photon energy respectively. The lifetime $\tau$ of the excitons in bulk ZnO is measured to be of several tens to hundreds picoseconds depending on temperature \cite{Travnikov}. Thus, two excitation regimes we used, nanosecond and femtosecond, are principally different. In the case of the excitation with ns pulses the creation of hot carriers is a kinetic process mainly determined by recombination of the carriers in a material and by the excitation power (eq. \ref{eq:rate}).
Finally, the nanocluster is heated by the accumulation of the excessive energy transfered from the new hot carriers created by absorption of photons.

Contrarily to the quasistationary carrier concentration under ns excitation, the fs laser pulses being much shorter than the exciton life-time create the population of hot electron-hole pairs which is independent of their time of life. In this case the carrier concentration will be determined only by the carrier generation rate ($G$). These carriers are present in the material only during their lifetime. Thus, at nearly the same free carrier concentration a fs pulse induces significantly lower heating than the nanosecond one. 
It should be noted that the heat losses by collisional, radiative and evaporative cooling of the nanoclusters are negligible for both ns and fs excitation regimes \cite{Lukyanchuk}.
Figure \ref{plfemto} presents typical photoluminescence spectra of free ZnO nanoclusters obtained after femtosecond excitation. The clusters were prepared at the same laser fluence of 3.5 J/cm$^{2}$ and at oxygen pressures of 6 and 8 mbar. The spectra have been taken two minutes after the ablation run.
The shape of the luminescence band appears 
to be independent on the gas pressure and it is very close to that observed 
in ZnO nanocrystalline films (Fig. \ref{plfemto}c). Both high and low-energy tails observed in fig. \ref{plfemto} are strongly reduced compared to the PL under nanosecond excitation.
This fact allows us to attribute the low-energy tail of the spectra taken under nanosecond excitation to heating of the nanoclusters during the laser pulse. We should note however that the reduced high energy tail of the excitonic spectra   is probably caused by the bandgap renormalization in the clusters with sizes below Bohr radius because the heating effect is reduced under fs excitation.
Absence of low-energy tail and the band maximum position at 3.28 eV also 
show that the majority of clusters reached the temperature of the ambient gas after their synthesis.  

\section{Conclusions}

We observed condensation and aggregation of ZnO nanoclusters prepared by pulsed laser ablation in gas phase. The nanoclusters suspended in the ambient oxygen gas are already crystallized and cooled down showing optical properties similar to those for solid films. Under nanosecond laser excitation the photoluminescence spectra of the interband recombination from the nanoclusters show the nanocluster heating by the accumulation of the energy transfered from the hot carriers. Under femtosecond laser excitation, when the thermal effect of laser irraditation is strongly reduced, the nanoclusters with sizes of several nanometers show a very efficient UV luminescence comparable to that of the high quality films.



\begin{thebibliography}{}

\bibitem{Bagnall} D.M. Bagnall, Y.F. Chen, Z. Zhu, T. Yao, S. Koyama, M.Y. Shen, and 
T. Goto, Appl. Phys. Lett. \textbf{70} (1997) 2230.

\bibitem{Tang} Z.K. Tang, G.K.L. Wong, P. Yu, M. Kawasaki, A. Ohtomo, H.Koinuma, Y. Segawa, Appl. Phys. Lett. \textbf{72} (1998) 3270.

\bibitem{Choopun}S. Choopun, R.D. Vispute, W. Noch, A. Balsamo, R.P. Sharma, T. Venkatesan, A. Iliadis, D.C. Look, Appl. Phys. Lett. \textbf{75} (1998) 3947.

\bibitem{Cao} H. Cao, Y.G. Zhao, X. Liu, E.W. Seelig, R.P.H. Chang, Appl. Phys.
Lett. \textbf{75} (2000) 1213.

\bibitem{Haga} K. Haga, M. Kamidaira, Y. Kashiwaba, T. Sekiguchi, H. Watanabe, J.Cryst. Growth \textbf{214--215} (2000) 77.

\bibitem{Yoon} K.H. Yoon, J.W. Choi, D.H. Lee, Thin Solid Films \textbf{302} (1997) 116.


\bibitem{Nakamura} T. Nakamura, H. Minoura, H. Muto, Thin Silid Films \textbf{405} (2002) 109.


\bibitem{ApSuSc2003} I. Ozerov, D. Nelson, A.V. Bulgakov, W. Marine, M. Sentis, 
Appl. Surf. Sci. \textbf{212--213} (2003) 349.

(arXiv: cond-mat/0311315)

\bibitem{Hartanto} A.B. Hartanto, X. Ning, Y. Nakata, T. Okada, Appl. Phys. A \textbf{78} (2004) 299.

\bibitem{Millon} E. Millon, O. Albert, J.C. Loulergue, J. Etchepare, D. Hulin, W. Seiler, J. Perri\`{e}re, J. Appl. Phys. \textbf{88} (2002) 6937.

\bibitem{Strategy} W. Marine, L. Patrone, B. Luk'yanchuk, M. Sentis, Appl. Surf. 
Sci. \textbf{154--155} (2000) 345.

\bibitem{Kawaguchi2002}Y. Kawaguchi, A. Narazaki, T. Sato, H. Niino, A. Yabe, Appl. Surf. Sci. 
\textbf{197--198} (2002) 268.

\bibitem{Claeyssens2002}F. Claeyssens, A. Cheesman, S.J. Henley, M. N.R. Ashfold, J. Appl. Phys. \textbf{92} (2002) 6886.



\bibitem{Bulgakov}A.V. Bulgakov, I. Ozerov, W. Marine, E-print (2003) http://arXiv.org/abs/physics/0311117.

\bibitem{Boufendi} L. Boufendi, J. Hermann, A. Bouchoule, B. Dubreuil, E. Stoffels, W.W.
Stoffels, M.L. de Giorgi, J. Appl. Phys. \textbf{76} (1994) 148.

\bibitem{JdePhys} I. Ozerov, A.V. Bulgakov, D. Nelson, R. Castell, M. Sentis, and W. Marine, J. Phys. IV \textbf{108} (2003) 37.

(arXiv: physics/0311053)

\bibitem{VanDijken} A. van Dijken, J. Makkinje, A. Meijerink, J. Lum. \textbf{92} (2001) 323.


\bibitem{Wang} L. Wang, N.C. Giles, J. Appl. Phys. \textbf{94} (2003) 973.

\bibitem{Yamamoto} A. Yamamoto, T. Kido, T. Goto, Y. Chen, T. Yao, Solid State Commun. 
\textbf{122}, 29 (2002).

\bibitem{Travnikov} V.V. Travnikov, A. Freiberg, S.F. Savikhin, J. Lum. \textbf{47} (1990) 107.

\bibitem{Lukyanchuk} B.S. Luk'yanchuk, W. Marine, Appl. Surf. Sci. \textbf{154--155} (2000) 314.

\end{thebibliography}
\end{document}